\documentclass[reprint]{revtex4-1}
\usepackage{graphicx}
\usepackage{amsmath}
\usepackage{amsfonts}
\usepackage{wasysym}
\usepackage{mathrsfs}
\usepackage{color}

\begin{document}
\title{Intermittency in generalized NLS equation with focusing six-wave interactions}
\author{D.S. Agafontsev$^{(a),(d)}$, V.E. Zakharov$^{(a),(b),(c),(d)}$}

\affiliation{\small \textit{ $^{(a)}$ P.P. Shirshov Institute of Oceanology, 36 Nakhimovsky prosp., Moscow 117218, Russia.\\
$^{(b)}$ Department of Mathematics, University of Arizona, Tucson, AZ, 857201, USA.\\
$^{(c)}$ P.N. Lebedev Physical Institute, 53 Leninsky ave., 119991 Moscow, Russia.\\
$^{(d)}$ Novosibirsk State University, 2 Pirogova, 630090 Novosibirsk, Russia.}}

\begin{abstract}
We study numerically the statistics of waves for generalized one-dimensional Nonlinear Schr{\"o}dinger (NLS) equation that takes into account focusing six-wave interactions, dumping and pumping terms. We demonstrate the universal behavior of this system for the region of parameters when six-wave interactions term affects significantly only the largest waves. In particular, in the statistically steady state of this system the probability density function (PDF) of wave amplitudes turns out to be strongly non-Rayleigh one for large waves, with characteristic ``fat tail'' decaying with amplitude $|\Psi|$ close to $\propto\exp(-\gamma|\Psi|)$, where $\gamma>0$ is constant. The corresponding non-Rayleigh addition to the PDF indicates strong intermittency, vanishes in the absence of six-wave interactions, and increases with six-wave coupling coefficient.
\end{abstract}

\pacs{02.30.Ik, 05.45.-a}

\maketitle


\section{Introduction}

The studies of statistics of waves for different nonlinear systems have drawn much scientific attention in the recent time \cite{mussot2009observation, genty2010collisions, taki2010third, hammani2010emergence, chung2011strong, onorato2013rogue}, especially since the first observation of optical rogue waves \cite{solli2007optical}. The experimental evidence of rogue waves emergence in both optics and hydrodynamics \cite{mussot2009observation, solli2007optical, kharif2003physical, dysthe2008oceanic, onorato2013rogue} imply that such waves may appear much more frequently than predicted by the approximation of random wave field governed by linear equations. Thus, the development of a consistent nonlinear theory for rogue waves phenomenon is urgently needed. 

Let us suppose, that wave field $\Psi$ is a random superposition of a multitude of uncorrelated linear waves. Then, the probability density function (PDF) of wave amplitudes in such state is Rayleigh one \cite{nazarenko2011wave},
\begin{equation}\label{Rayleigh0}
\mathcal{P}_{R}(|\Psi|) = \frac{2|\Psi|}{\sigma^{2}}e^{-|\Psi|^{2}/\sigma^{2}}.
\end{equation}
Here $\sigma$ is constant and we use normalization for the PDF as $\int\mathcal{P}(|\Psi|)\,d |\Psi|=1$. Below it will be convenient for us to study the PDFs for normalized squared amplitudes $I=|\Psi|^{2}/\langle|\Psi|^{2}\rangle$, where $\langle|\Psi|^{2}\rangle$ is the averaged squared amplitude. The quantity $I$ has the meaning of ``relative power''; values $I\ll 1$ represent small waves, $I\sim 1$ -- moderate waves and $I\gg 1$ -- large waves. Then, Rayleigh PDF (\ref{Rayleigh0}) takes the simple form 
\begin{equation}\label{Rayleigh}
\mathcal{P}_{R}(I) = e^{-I}.
\end{equation}
Here we used relations $\int 2x F(x)\,dx=\int F(x)\,dx^{2}$ and $\langle|\Psi|^{2}\rangle=\sigma^{2}$, the latter of which follows from Eq. (\ref{Rayleigh0}). We will call the PDF (\ref{Rayleigh}) as Rayleigh one for simplicity. 

If evolution of a random wave field is governed by linear equations, then the superposition of linear waves stays uncorrelated, and the PDF remains Rayleigh one (\ref{Rayleigh}). The nonlinear evolution may introduce correlation, that in turn may lead to enhanced appearance of large waves. Today the most popular nonlinear model for the description of rogue waves in both optics and hydrodynamics is the focusing one-dimensional Nonlinear Schr{\"o}dinger (NLS) equation \cite{kharif2003physical, dysthe2008oceanic, onorato2013rogue},
\begin{equation}\label{nlse}
i\Psi_t + \Psi_{xx} - \Psi + |\Psi|^2 \Psi = 0.
\end{equation}
Here $t$ is time, $x$ is spatial coordinate and $\Psi$ is wave field or wave field envelope. The simplest ``condensate'' solution of this equation $\Psi=1$ is modulationally unstable. The development of this instability from initially small perturbation $\epsilon(x)$,
\begin{equation}\label{condensate}
\Psi|_{t=0} = 1+\epsilon(x),\quad |\epsilon(x)|\ll 1,
\end{equation}
leads to formation of ``integrable turbulence'' \cite{zakharov2009turbulence}, during which rogue waves may emerge. Some of the rogue wave solutions can by analyzed analytically, see \cite{kuznetsov1977solitons, peregrine1983water, akhmediev1987exact, zakharov2011soliton, zakharov2013nonlinear}.

However, as was demonstrated in \cite{agafontsev2014integrable}, in the framework of the modulation instability (MI) model the PDF of wave amplitudes, averaged over realizations of initial perturbations, does not exceed significantly Rayleigh distribution (\ref{Rayleigh}). The development of the MI leads after a very long evolution to the asymptotic state of the stationary integrable turbulence. The PDF in this state coincides with Rayleigh one (\ref{Rayleigh}). During the evolution toward the asymptotic state, the PDF may significantly deviate from (\ref{Rayleigh}), however, it does not exceed Rayleigh PDF by more than 3 times.

In \cite{walczak2014optical} the different scenario for the integrable turbulence in the framework of the focusing NLS equation was studied, with the incoherent wave field as initial conditions. In this case the system after a short evolution reaches the stationary state, where the tail of the PDF at large amplitudes may exceed Rayleigh distribution (\ref{Rayleigh}) by orders of magnitude. However, this tail still decays according to modified Rayleigh law $\propto e^{-\gamma I}$, where $\gamma>0$ is constant.

In the current study we demonstrate how even sufficiently small additional terms to the focusing NLS equation may qualitatively change the behavior of the PDF at large amplitudes. The natural way to modify Eq. (\ref{nlse}) consists in addition to it's left-hand side the next-order nonlinear terms that appear in the perturbation theory expansion beyond the four-wave interactions term $|\Psi|^{2}\Psi$. One of these terms is the six-wave interactions $\alpha|\Psi|^{4}\Psi$, that appears in many physical models including Bose-Einstein condensation \cite{pitaevskii2003condensation, brazhnyi2006dark}, surface water waves \cite{agafontsev2007deep, agafontsev2008bifurcations}, pattern formation in the framework of the Ginzburg -- Landau equation \cite{cross1993pattern}, dissipative solitons in lasers \cite{soto2000pulsating}, optical fibers \cite{gabitov2002nonlinearity}, and so on. Here $\alpha$ is the six-wave coupling coefficient. In this publication we 
consider only the focusing case $\alpha>0$.

However, addition of only the focusing six-wave interactions to Eq. (\ref{nlse}) results in generation of blow-up collapses in a finite time, even for very small coefficients $\alpha$. Indeed, as we checked numerically, the MI generates quasi-solitons that collide inelastically in the presence of six-wave interactions. Similarly to non-collapsing models \cite{jordan2000self}, during these collisions the larger quasi-solitons increase while the smaller quasi-solitons decay. After some time one of the quasi-solitons becomes so large, that the influence of six-wave interactions for this wave exceeds significantly the four-wave interactions, $\alpha|\Psi|^{4}\gg |\Psi|^{2}$. Then, it's subsequent evolution can be effectively described by the quintic NLS equation that generates blow-up collapses in a finite time \cite{zakharov1972collapse, dyachenko1992optical, chung2011strong}.

In order to regularize these collapses we add to the left-hand side of Eq. (\ref{nlse}) small linear and nonlinear dumping terms $-id_{l}\Psi_{xx}$ and $id_{3p}|\Psi|^{4}\Psi$ respectively, where $d_{l}>0$ and $d_{3p}>0$, $d_{3p}\ll\alpha$, are small constants. The first term prevents the appearance of too large gradients and may appear in the context of the complex Ginzburg -- Landau equation, modeling viscosity in hydrodynamics, as well as filtering and gain dispersion in optics \cite{sakaguchi2001stable, falkovich2004non, malomed2007solitary, bale2010intracavity}. The second term prevents the appearance of too large amplitudes and may appear from three-photon absorption \cite{boyd2003nonlinear} in optics and four-body collisions in Bose-Einstein condensates \cite{ferlaino2009evidence}. In order to balance the system we also add small deterministic pumping term $ip\Psi$ to the right-hand side of Eq. (\ref{nlse}). Here $p>0$ is small constant. We checked other statistically homogeneous in x-space 
pumping terms including chaotic forcing \cite{chung2011strong}, and also other nonlinear dumping terms including two-photon absorption, and found no significant difference in our results. Thus, 
we come to the following generalized NLS equation:
\begin{eqnarray}
&& i\Psi_t + (1-id_{l})\Psi_{xx} - \Psi + |\Psi|^2 \Psi + \nonumber\\ 
&& + (\alpha + id_{3p})|\Psi|^4 \Psi = ip\Psi. \label{GL5}
\end{eqnarray}

As we observe in our simulations, if the coefficients before the additional terms are sufficiently small, $d_{l}, d_{3p}, p, \alpha\ll 1$, then the initial evolution of Eq. (\ref{GL5}) is very similar to that of the focusing NLS equation (\ref{nlse}) for both the MI development and the incoherent wave field initial conditions. Then, the system (\ref{GL5}) deviates from Eq. (\ref{nlse}) and gradually approaches to the statistically steady state, which has independent on time statistical characteristics. As we checked, both types of initial conditions lead to the same statistically steady state (see, e.g., FIG.~\ref{fig:fig1}a, and also \cite{chung2011strong}).

The statistically steady state is the energy equilibrium of Eq. (\ref{GL5}), since the energy input due to the pumping term is statistically balanced by the energy output due to the dumping terms. It turns out that this equilibrium is very sensitive with respect to the specific values of the coefficients $(d_{l},d_{3p},p,\alpha)$, even when these coefficients are all small. Thus, for different sets of these coefficients it is possible to get statistically steady states with large influence of six-wave interactions already for medium waves $I\sim 1$, or with small four- and six-wave interactions even for the largest waves $I\gg 1$ present in the system. Nevertheless, as we report in this study, in the statistically steady state of Eq. (\ref{GL5}) the PDF of wave amplitudes has universal behavior.

Thus, if the six-wave interactions term turns out to be large already for medium waves, then the system can be effectively described by the modified quintic NLS equation. As was already demonstrated in \cite{chung2011strong}, in this case collapses appear randomly is space and time, and their evolution and subsequent regularization are self-similar. The PDF of wave amplitudes is close then to Rayleigh PDF for small and moderate waves, but has a ``fat tail'' that decays as $\propto |\Psi|^{-8}$ at large amplitudes.

In the current publication we study numerically the different scenario, when the six-wave interactions term affects significantly only the largest waves. We demonstrate that in this case the PDF $\mathcal{P}(I)$ is close to Rayleigh one for small and moderate waves, while it's tail at large amplitudes $I\gg 1$ decays close to $\propto e^{-\gamma \sqrt{I}}$, where $\gamma>0$ is constant. The corresponding non-Rayleigh addition to the PDF vanishes in the absence of six-wave interactions and increases with six-wave coupling coefficient $\alpha$ as $\gamma\propto\alpha^{1/2}$. The typical rogue wave events in this system have short duration in time, and represent in space a singular high peak resembling the modified Peregrine solution \cite{peregrine1983water} of the focusing NLS equation (\ref{nlse}).

The paper is organized as follows. In the next section we describe the numerical methods that we used in the framework of the current study. Then we present the results of our numerical experiments. The final section contain conclusions and acknowledgements.


\section{Numerical methods}

We integrate Eq. (\ref{GL5}) numerically in the box $x\in[-L/2, L/2]$, $L=64\pi$, with periodic boundary. In contrast to the MI development for the focusing NLS equation (\ref{nlse}), we do not observe the concentration of wave action $N = \int |\Psi|^{2}\, dx$ into modes with extremely large scales \cite{agafontsev2014integrable}. This allows us to perform simulations using not very long computational boxes $L$; we checked that boxes longer than $L=64\pi$ do not provide us different results. We implement the same method of numerical simulations that was developed in \cite{agafontsev2014integrable}. Specifically, we use Runge-Kutta 4th-order method with adaptive change of the spatial grid size $\Delta x$ and Fourier interpolation of the solution between the grids. In order to prevent appearance of numerical instabilities, time step $\Delta t$ 
changes with $\Delta x$ as $\Delta t = h\Delta x^{2}$, $h \le 0.1$.

We start simulations on the grid with $M=4096$ nodes. We average our results across 1000 random realizations of initial data. The choice of the initial data is explained below. For most of our simulations we arrive to the statistically steady state before time $t=300$. Since the characteristics of this state do not depend on time, we additionally average these results over interval $t\in[350, 400]$. Also, there were two simulations when the statistically steady state was reached near $t=1000$. In these cases that will be outlined below, we average the results over interval $t\in[950, 1000]$. We checked the results against the size of the ensembles, the parameters of our numerical scheme and the implementation of other numerical methods (Runge-Kutta 5th-order, Split-Step 2nd- and 4th-order methods \cite{muslu2005higher, mclachlan1995numerical}), and found no difference. 

\begin{figure}[t] \centering
\includegraphics[width=8.5cm]{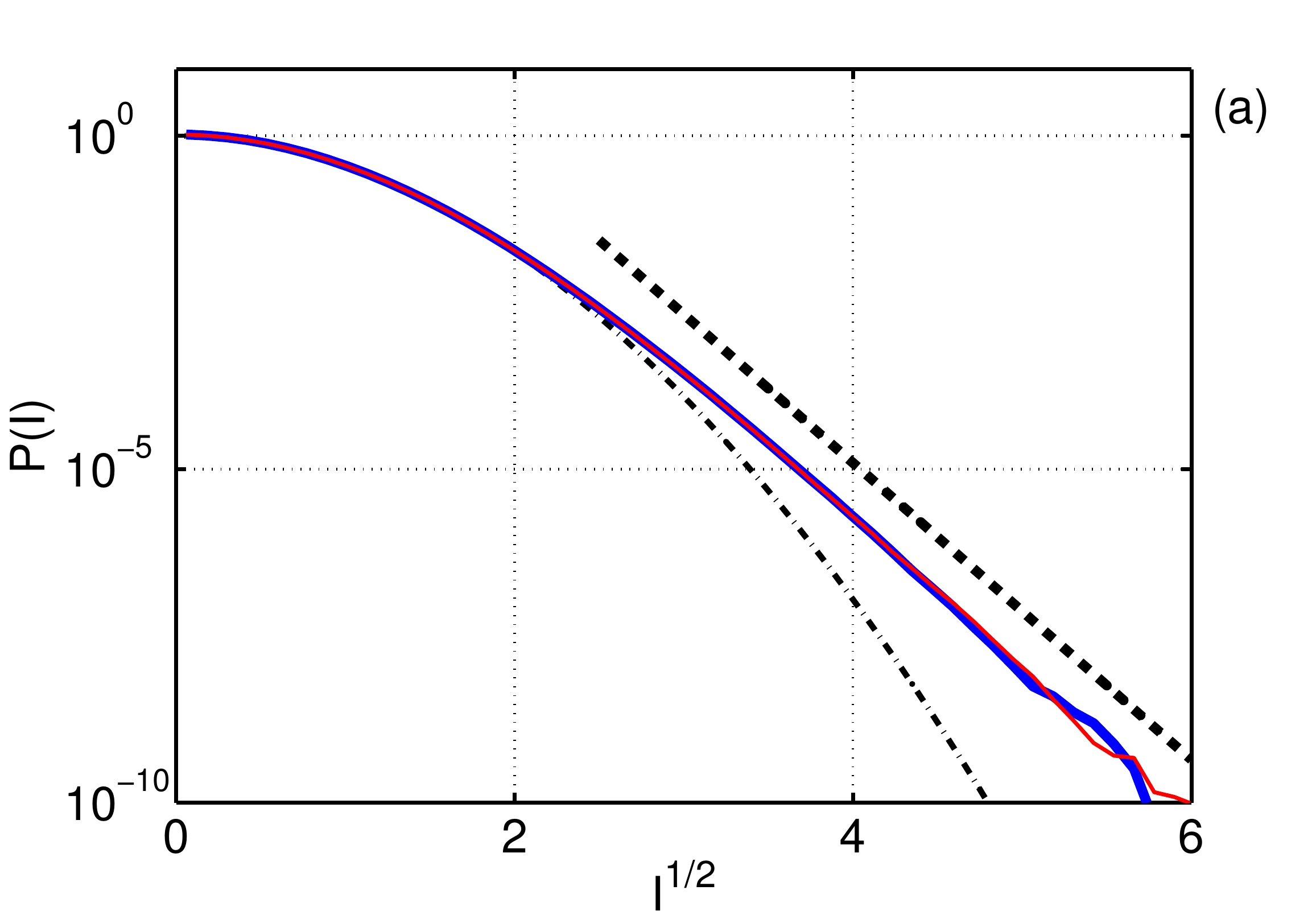}
\includegraphics[width=8.5cm]{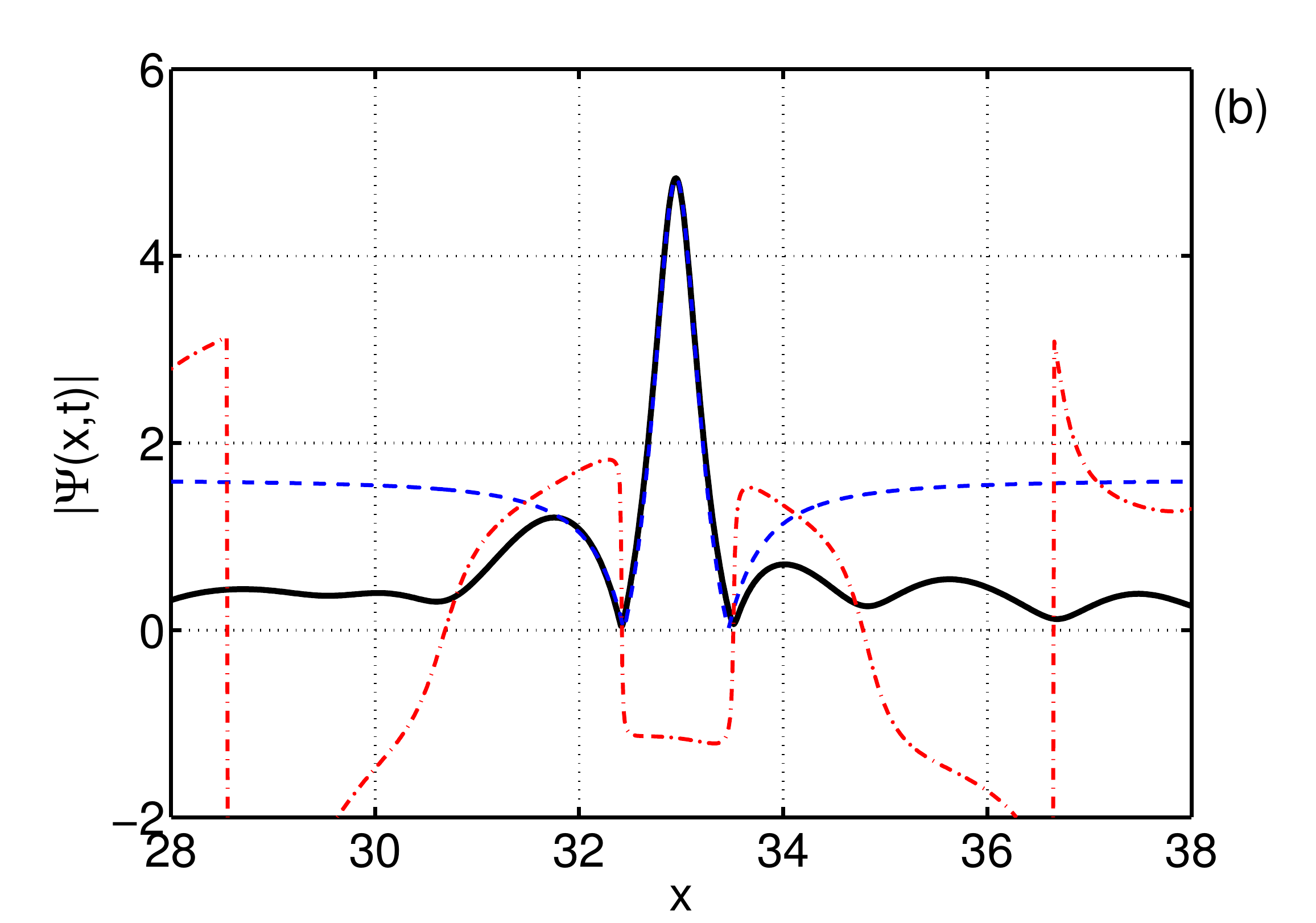}

\caption{\small {\it (Color on-line) (a) The PDFs $\mathcal{P}(I)$ in the statistically steady state, versus $\sqrt{I}$. Bold blue line corresponds to the incoherent wave field initial conditions (\ref{incoherent_WF}) with perturbation parameters $A_{0}=1$, $\theta=1$; thin red line -- to the MI development (\ref{condensate}) with $A_{0}=10^{-5}$, $\theta=5$; see Eq. (\ref{noise}). Thick dashed line indicates decay law $\propto e^{-\gamma\sqrt{I}}$ with $\gamma\approx 5.13$, thin dash-dot line is Rayleigh PDF (\ref{Rayleigh}). (b) Spatial distribution $|\Psi(x)|$ for a typical rogue wave at the statistically steady state. This rogue wave event had duration in time of about $\Delta T\sim 0.5$, and reached maximum amplitude $\max|\Psi|\approx 4.83$ at time $t\approx 379$; the statistically steady state has average squared amplitude $\langle|\Psi|^{2}\rangle\approx 0.7$. Dashed blue line is fit by function (\ref{Peregrine}) with $A\approx -1.6$, $b\approx 2.4$, $x_{0}\approx 32.9$, dash-dot red 
line is argument (phase) $\mathrm{arg}\,\Psi$. 
The parameters of Eq. (\ref{GL5}) for figures (a) and (b) are: $\alpha=1/25$, $d_{l}=10^{-3}$, $d_{3p}=10^{-3}$, $p=5\times 10^{-3}$.}}
\label{fig:fig1}
\end{figure}

We performed numerical experiments starting from the initial data that corresponds to the problems of (a) the MI development (\ref{condensate}), and (b) the propagation of incoherent wave field 
\begin{equation}\label{incoherent_WF}
\Psi|_{t=0}=\epsilon(x),
\end{equation}
when random perturbation $\epsilon(x)$ is not necessarily small. We found that the final statistically steady states for these problems coincide (see, e.g., FIG.~\ref{fig:fig1}a, and also \cite{chung2011strong}), irrespective of the different types of the perturbations $\epsilon(x)$ that we studied. This fact demonstrates that the statistically steady state of Eq. (\ref{GL5}) does not depend on the initial conditions $\Psi|_{t=0}$. 

In our final numerical experiments we used statistically homogeneous in space perturbations $\epsilon(x)$ in the form 
\begin{equation}\label{noise}
\epsilon(x)=A_{0}\bigg(\frac{\sqrt{8\pi}}{\theta L}\bigg)^{1/2} \sum_{k}e^{-k^{2}/\theta^{2}+i\xi_{k}+ikx},
\end{equation}
for both of the problems (\ref{condensate}) and (\ref{incoherent_WF}). Here $k=2\pi n/L$ is wavenumber, $n$ is integer, $A_{0}$ is perturbation amplitude, $\theta$ is perturbation width in k-space, and $\xi_{k}$ are arbitrary phases for each $k$ and each perturbation realization within the ensemble of initial data. As was shown in \cite{agafontsev2014integrable}, the average squared amplitude of perturbation (\ref{noise}) in x-space is very close to $A_{0}^{2}$, 
\begin{eqnarray}
\overline{|\epsilon|^{2}} = \frac{1}{L}\int_{-L/2}^{L/2} |\epsilon(x)|^{2}\, dx \approx A_{0}^{2}.\label{noise_amplitude}
\end{eqnarray}

Since the statistically steady state does not depend on the initial conditions, and for the problem (\ref{incoherent_WF}) we arrive to this state significantly faster, we performed most of our numerical experiments starting from the incoherent wave field (\ref{incoherent_WF}). Technically, we used perturbation parameters $A_{0}=1$ and $\theta=1$; however, we checked that the statistically steady states do not depend on the specific values of these parameters. 

In the next Section we concentrate on our numerical experiments with parameters $d_{l}=10^{-3}$, $d_{3p}=10^{-3}$, $p=5\times 10^{-3}$ and $\alpha=1/25$. In this case the statistically steady state realizes with the average squared amplitude equal to $\langle |\Psi|^{2}\rangle\approx 0.7$. Thus, for moderate waves $|\Psi|\sim 1$ the influence of six-wave interactions term $\alpha|\Psi|^{4}\Psi$ turns out to be small in comparison with the four-wave interactions $|\Psi|^{2}\Psi$, since $\alpha|\Psi|^{4}\sim 1/25$ and $|\Psi|^{2}\sim 1$. However, already for $|\Psi|\sim 3$ these interactions become comparable, as $\alpha|\Psi|^{4}\sim 3$ and $|\Psi|^{2}\sim 9$.

Below we also demonstrate the dependence of our results on six-wave coupling coefficient $\alpha$. In addition to these experiments, we studied several different sets of dumping and pumping coefficients $d_{l}$, $d_{3p}$ and $p$. We found that if the four coefficients $(d_{l},d_{3p},p,\alpha)$ provided such statistically steady states, for which the six-wave interactions term affected significantly only the largest waves, we always came to the same qualitative conclusions. Note that the scaling and gauge transformations 
$$
x=\xi/A, \quad t=\tau/A^{2}, \quad \Psi=\Phi\times A e^{i(1-1/A^{2})\tau},
$$
where $A>0$ is constant, do not change the form of Eq. (\ref{GL5}), 
\begin{eqnarray}
&& i\Phi_{\tau} + (1-id_{l})\Phi_{\xi\xi} - \Phi + |\Phi|^2 \Phi + \nonumber\\ 
&& + A^{2}(\alpha + id_{3p})|\Phi|^4 \Phi = i\frac{p}{A^{2}}\Phi. \nonumber
\end{eqnarray}
This means that the sets of coefficients $(d_{l},d_{3p},p,\alpha)$ and $(d_{l},A^{2}d_{3p},p/A^{2},A^{2}\alpha)$ provide equivalent statistically steady states with average squared amplitudes relating as $\langle|\Psi|^{2}\rangle = A^{2}\langle|\Phi|^{2}\rangle$, that in turn leads to the same PDFs for normalized squared amplitudes $\mathcal{P}(I)$.

\begin{figure}[t] \centering
\includegraphics[width=8.5cm]{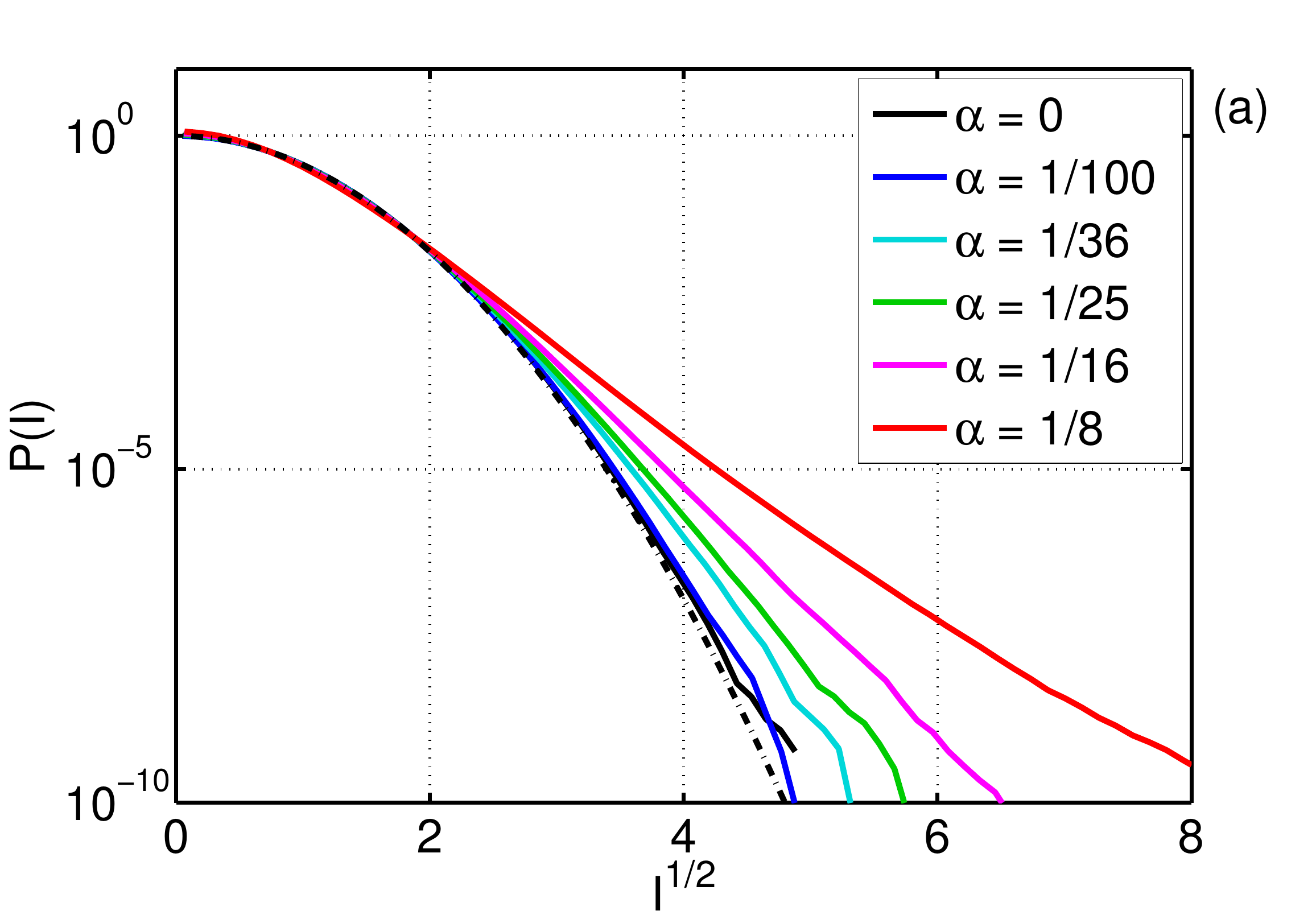}
\includegraphics[width=8.5cm]{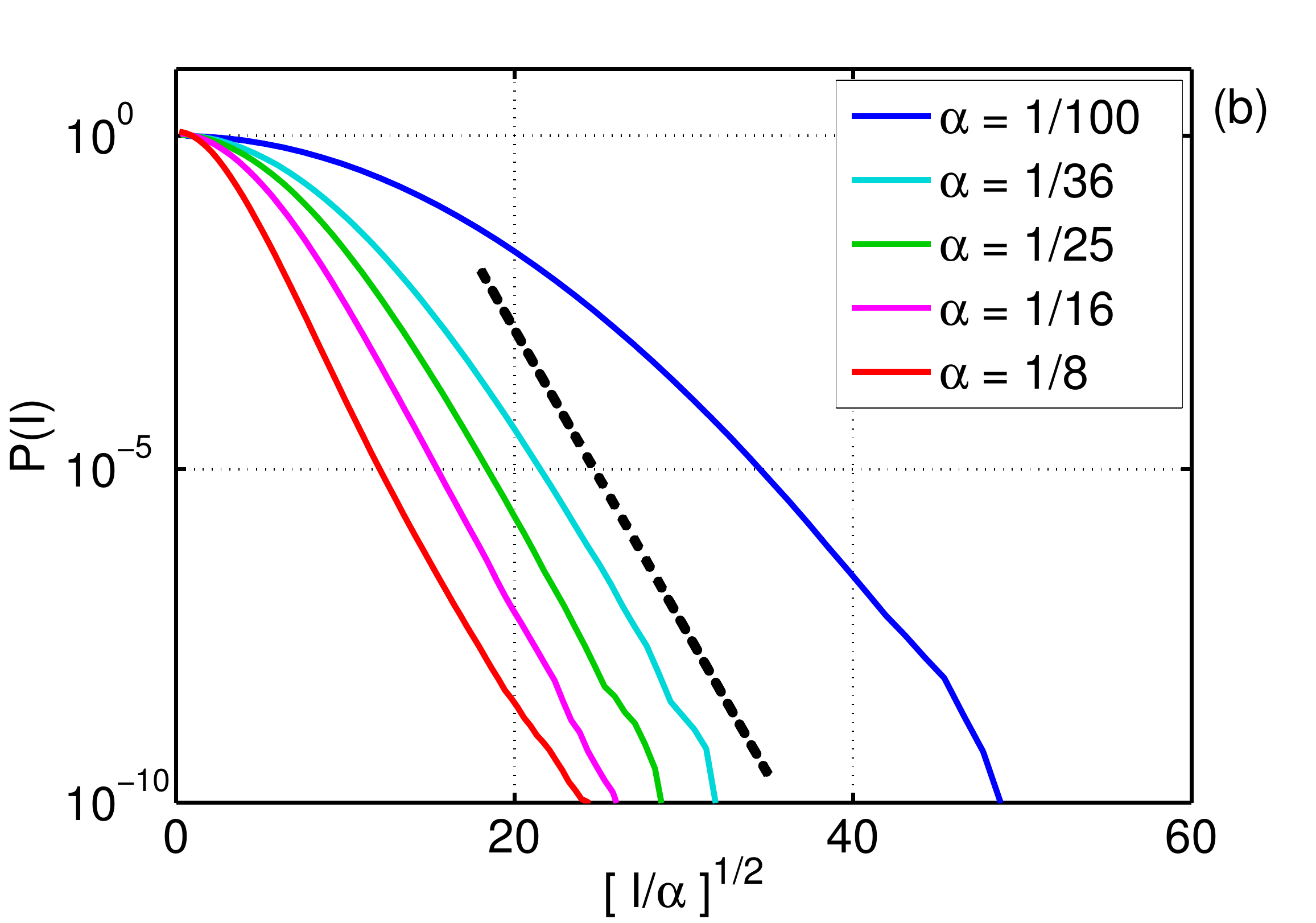}

\caption{\small {\it (Color on-line) The PDFs $\mathcal{P}(I)$ in the statistically steady states at different $\alpha$, versus $\sqrt{I}$ (a) and $\sqrt{I/\alpha}$ (b). Dumping and pumping coefficients are fixed to $d_{l}=10^{-3}$, $d_{3p}=10^{-3}$, $p=5\times 10^{-3}$. The initial conditions are the incoherent wave field (\ref{incoherent_WF}) with perturbation parameters $A_{0}=1$, $\theta=1$. Thin dash-dot line on figure (a) is Rayleigh PDF (\ref{Rayleigh}), thick dashed line on figure (b) indicates decay law $\propto e^{-\tilde{\gamma}\sqrt{I/\alpha}}$ with $\tilde{\gamma}\approx 1.03$. Experiments with $\alpha=0$ and $\alpha=1/100$ were carried out up to final time $t=1000$, with averaging of the results over interval $t\in[950,1000]$. }}
\label{fig:fig2}
\end{figure}


\section{Waves statistics in the statistically steady state}

FIG.~\ref{fig:fig1}a demonstrates the PDF $\mathcal{P}(I)$ at the statistically steady state of Eq. (\ref{GL5}) with parameters $d_{l}=10^{-3}$, $d_{3p}=10^{-3}$, $p=5\times 10^{-3}$, and $\alpha=1/25$. At small and moderate waves $I\le 2$ the PDF turns out to be close to Rayleigh one (\ref{Rayleigh}). However, for larger waves $I\gtrsim 3$ the PDF decays as $\propto e^{-\gamma \sqrt{I}}$, thus significantly deviating from Rayleigh one. For example, at $I=5$ the PDF is about $10^{3}$ times larger than Rayleigh distribution (\ref{Rayleigh}). 

The typical rogue waves, that contribute to the non-Rayleigh tail of the PDF, are rare events that have short duration in time and resemble in space the modified Peregrine solution \cite{peregrine1983water} of the focusing NLS equation (see FIG.~\ref{fig:fig1}b),
\begin{eqnarray}
\Psi(x)\approx A\bigg[1 - \frac{4}{1+2b^{2}(x-x_{0})^{2}} \bigg]. \label{Peregrine}
\end{eqnarray}
Here $A$, $b$ are constants, and $x_{0}$ is the position of the maximum amplitude. Note that the argument (phase) $\mathrm{arg}\,\Psi(x)$ of the rogue wave is almost constant near the amplitude maximum, as is the case for anzats (\ref{Peregrine}). However, critical distinctions between the rogue waves of Eq. (\ref{GL5}) and the Peregrine solution are present. First, the coefficients $A$ and $b$ are significantly different from unity (see FIG.~\ref{fig:fig1}b), which would be the case for the Peregrine solution. Second, the rogue waves of Eq. (\ref{GL5}) appear in the dissipative system, and, as we checked numerically, at the time of their appearance the dissipation is enhanced (see also \cite{chung2011strong}).

The non-Rayleigh addition to the PDF practically vanishes in the absence of six-wave interactions $\alpha=0$, and increases with $\alpha$, FIG.~\ref{fig:fig2}a. In order to qualitatively study this dependence, we represent the PDFs $\mathcal{P}(I)$ on FIG.~\ref{fig:fig2}b versus $\sqrt{I/\alpha} = |\Psi|/\sqrt{\alpha\langle|\Psi|^{2}\rangle}$. In the physical sense this renormalization choice is natural, since as follows from Eq. (\ref{GL5}), the quantity $\alpha\langle|\Psi|^{2}\rangle$, being dimensionless, describes the ratio between the six-wave and the four-wave interactions. FIG.~\ref{fig:fig2}b shows that if the six-wave coupling coefficient $\alpha$ is not very close to zero, then the PDFs decay versus $\sqrt{I/\alpha}$ with almost the same slope. This means that the slope $\gamma$ of the decay $\mathcal{P}(I)\propto e^{-\gamma \sqrt{I}}$ depends on the six-wave coupling coefficient $\alpha$ as 
\begin{eqnarray}
\gamma\approx \tilde{\gamma}\,\alpha^{1/2},\label{gamma_alpha}
\end{eqnarray}
where $\tilde{\gamma}$ is constant determined by dumping and pumping parameters. The deviations from this relation are seen for very small $\alpha$ (see, e.g., $\alpha=1/100$), when the six-wave interactions term is small even for the largest waves $I\gg 1$ and the entire PDF is very close to Rayleigh distribution (\ref{Rayleigh}), and also for sufficiently large $\alpha$ (see, e.g., $\alpha=1/8$, at large amplitudes), where the six-wave and the four-wave interactions become comparable.


\section{Conclusions and acknowledgements}

In the current publication we studied the statistics of waves for generalized one-dimensional NLS equation (\ref{GL5}) that takes into account focusing six-wave interactions, dumping and pumping terms. Our main motivation was to investigate whether sufficiently small additional terms to the focusing NLS equation may qualitatively change the behavior of the PDF at large amplitudes. 

We found that the system (\ref{GL5}) approaches with time to the statistically steady state that has statistical characteristics independent on time and initial conditions. This state, being the energy equilibrium of Eq. (\ref{GL5}), is determined only by the set of coefficients $(d_{l},d_{3p},p,\alpha)$. We demonstrated the universal behavior for those statistically steady states, where the six-wave interactions term affects significantly only the largest waves $I\gg 1$. Namely, in such states the PDF $\mathcal{P}(I)$ turns out to be close to Rayleigh one (\ref{Rayleigh}) for small and moderate waves $I\lesssim 1$, and decays as $\propto e^{-\gamma\sqrt{I}}$ at large waves $I\gg 1$. The corresponding non-Rayleigh addition to the PDF vanishes in the absence of six-wave interactions $\alpha=0$, and increases with $\alpha$ as $\gamma\propto\alpha^{1/2}$. The non-Rayleigh tail of the PDF indicates strong intermittency in the system, as it is created by the appearance of rare rogue waves of short duration in 
time. These rogue waves represent in space a singular high peak (\ref{Peregrine}) resembling the modified Peregrine solution.

The authors thank E. Kuznetsov for valuable discussions concerning this publication, M. Fedoruk for access to and V. Kalyuzhny for assistance with Novosibirsk Supercomputer Center. This work was done in the framework of RSF grant 14-22-00174.

\end{document}